\documentstyle[12pt]{article}

\newcommand{\aaa}{{\cal A}}

\newcommand{\be}[1]{\begin{equation}\label{#1}}
\newcommand{\bra}[1]{{\langle #1 |}}
\newcommand{\braket}[2]{{\langle #1 |\, #2\rangle}}

\newcommand{\closure}{{\rm\bf Cl}}

\newcommand{\degree}{{\rm deg}}

\newcommand{\ee}{\end{equation}}

\newcommand{\fff}{{\cal F}} 

\newcommand{\ket}[1]{{|\, #1\rangle}}
\newcommand{\ketbra}[2]{{|\, #1\rangle\langle #2|}}
\newcommand{\kkk}{{\cal K}}

\newcommand{\mmm}{{\cal M}} 
\newcommand{\om}{\omega}
\newcommand{\omg}{\Omega}
\newcommand{\ppp}{{P}} 
\newcommand{\qqq}{{Q}} 
\newcommand{\rrr}{{\cal R}}

\newcommand{\spn}{{\rm span}}

\newcommand{\tttt}{{T}} 
\newcommand{\uuu}{{\cal U}} 

\title{
Quantization of discretized spacetimes and the correspondence
principle }

\author{
Ioannis Raptis\footnote{Algebra and Geometry Section,
    Department of Mathematics, University of Athens,
    Panepistimioupolis 157 84, Greece, e-mail:
    iraptis@eudoxos.dm.uoa.gr}
and Roman R. Zapatrin\footnote{Friedmann
    Laboratory for Theoretical Physics, SPb UEF, Griboyedova, 30-32,
    191023, St.Petersburg, Russia, e-mail: gudrs@mail.ru}
}

\date{}

\begin{document}

\maketitle

\begin{abstract}
An algebraic quantization procedure for discretized spacetime
models is suggested based on the duality between finitary
substitutes and their incidence algebras. The provided limiting
procedure that yields conventional manifold characteristics of
spacetime structures is interpreted in the algebraic quantum
framework as a correspondence principle.
\end{abstract}

\section*{Motivation}

Current physical theory predicts that at small scales the
conventional picture of spacetime as a 4-dimensional differential
manifold breaks down to something more discrete, finitary and
quantum. This inadequacy of the smooth spacetime manifold is on one
hand due to the ideal character of event determinations of a
classical observer, on the other due to the appearance of
singularities.

To deal with the first shortcoming of the manifold model,
we insist that realistic models of measurement should be pragmatic:
we actually perform a finite number of observations and record a
finite number of events.  Thus, the conventional infinitude of
events that we adopt to model classical spacetime structure seems
to be a gross generalization of little operational value: we have
no actual experience of a continuous infinity of events and
their infinitesimal differential separation can not be recorded
in the laboratory.

However, due to the success of the classical model of observation
at large scales one expects a connection between the
realistic models and their ideal counterpart. The anticipated
connection could be formulated in terms of a correspondence
principle. That is to say, the structure of ideal observations
arises in some kind of limit of the structures of pragmatic
measurements. The aim of this paper is to provide a physical
account for this correspondence.

The central object of the correspondence will be the {\em classical
event} living in the limit manifold. This is modelled by a point
there and a classical observer records no quantum interference of
events. But in the pragmatic regime we interpret the event as a
{\em pure state of spacetime} and we admit coherent quantum
superposition between events. In this sense
the quantum substratum of pragmatic events 'decoheres' to the classical
point event of the limit manifold and the physical meaning of
the correspondence principle is the usual quantum one due to
Bohr.

We use the finitary substitutes proposed by Sorkin (1991) to model
combinatorial relations between events in realistic measurements.
The incidence algebras due to Rota (1968) are employed to
accomplish the same thing but operationally, that is,
algebraically. There is a duality implicit here that is pregnant to
familiar notions about the duality of quantum dynamics. In our
treatment the Sorkin model is held to represent an evolution of
states much like the Schr\"odinger picture of quantum systems,
while the Rota model recalls the evolution of operators similar to
the Heisenberg picture. Our approach lies with
the latter picture although one is always able to switch
back to the Sorkin scheme (Zapatrin, 1998). In brief, we propose
that posets describe the dynamical evolution of events when the
algebras describe the dynamical evolution of event determinations
(operations).

Our algebraic approach is constructive, that is, we provide a
matrix representation for the algebras employed. The latter possess
preferred elements that represent the pragmatic observations that
in the ideal limit are expected to yield the irreducible elements
of classical observations: the manifold point events. They
constitute abelian subalgebras of the incidence algebras and are
coined {\em stationaries}. Of course, actual pragmatic observations
are expected to effect dynamical transitions between quantum states
of spacetime (stationaries). These are modelled by non-commuting
operations in the algebra of pragmatic events and are called {\em
transients}.  We anticipate that stationaries will correspond to
classical point events in the limit manifold (in Section \ref{s4} we
show how sequences of station{\it a}ries become station{\it e}ry
recording classical events), while transients to some kind of tangent
structure at an event.

To deal with the second shortcoming of the manifold model we note
that at the pragmatic level of observations there are no
points, but only algebras. We call this feature {\em
alocality}.  Nevertheless, we require the classical correspondence
limit to yield the familiar local structure of spacetime: the point
event and the space tangent to it. Since any point of the limit
manifold can be the host of a singularity of some important
physical field, the alocal quantum pragmatic substratum may prove
to be an effective resolution of spacetime into finite quantum
elements. The quantum substratum is asingular (finite) because
alocal. Thus, from our perspective, locality, that is, the
assumption of a differential continuum for spacetime (Einstein,
1924), is the prime reason for
singularities, so that at the pragmatic level of observation we
abandon it. We suggest a plausible quantum theory of spacetime
structure with a strong operational and finitistic character.
Then, based on the algebraic models of pragmatic observations we
may develop a non-commmutive differential geometry to erect a
quantum theory of gravity on it (Parfionov and Zapatrin, 1995).
The correspondence limit suggested in the present paper may also
be employed in this context to recover the classical algebra of
spacetime observables and the conventional differential geometry
of the spacetime manifold (used to describe general relativity)
from the pragmatic non-commutative quantum substratum.

It should be mentioned that our algebraic approach is rather
flexible in the following sense: alternatively to the novel notion
of alocality in the pragmatic regime we can formulate
the notion of {\em nearest neighbour} connections. The
latter was assumed by Finkelstein (1985)
to be the principal characteristic of the physical causal topology in
the quantum deep so as to localise in some sense a causality relation
between events (Bombelli {\em et al.}, 1987). This causality relation
was modelled by a partial order. Thus, if we physically interpret
Sorkin's finitary substitutes as causal sets (Sorkin, 1995), a recent
result (Zapatrin, 1998) allows us to represent the 'nearest neighbour'
causal connection between events algebraically in the pragmatic
regime thus vindicate Finkelstein's demand for an algebraic
representation of immediate causality (Finkelstein, 1985). There, in
turn, we have the advantage of interpreting this connection
operationally and study its quantum properties. The question we are
confronted with is:  what is the physical meaning in the pragmatic
algebraic regime of the Sorkin--Finkelstein local causality? As an
answer we expect a formulation of local causality in operational terms
with a quantum interpretation, something that is missing in Sorkin's
picture which is dual to ours.  Affine to this question is the
following one: how our pragmatic event determinations accord with and
be adequate models of the causal structure of the world at small
scales?

Finally, inspired by the Sorkin (1991) approach, we contend that
pragmatic measurements can be subjected to refinements. In passing
to the dual picture we deal with algebras and, in accordance with
the correspondence limit, the ideal ultimate refinement corresponds
to what is known as the algebra of classical observables
(coordinates) and the manifold supporting them.

\section{Finitary preliminaries}\label{s1}

A finitary topological space is defined in (Sorkin, 1991) as a
space with any bounded region in it consisting of a finite number
of points. This seems to be a reasonable model for actual
measurements involving a finite number of events during experiments
of finite spatiotemporal extent.

Any finitary topological space $\mmm$ can be equivalently pictured as
poset. Introduce the relation "$\to$" between points of $\mmm$

\[
p\to q \quad \Leftrightarrow \quad \mbox{ the constant sequence }
\{p,p,\ldots,p,\ldots\} \quad \mbox{ tends to } q
\]

\noindent using the standard definition of convergence: a sequence
$\{p_1,p_2,\ldots \}\to q$ iff for any open set $U$ containing $q$
there is a number $N_U$ such that $p_n\in U$ for any $n\ge N_U$.

The obtained relation "$\to$" is always reflexive ($p\to p$) and
transitive ($p\to q$, $q\to r$ imply $p\to r$). Vice versa, any
quasiordered set $(\mmm, \to)$ acquires a topology defined through the
closure operator on subsets $P\subseteq M$:

\[
\closure P =
\{q\,:\, \exists p \in P \quad p\to q \}
\]

For technical reasons (see Section \ref{s2}) we employ the
Alexandrov (1956) construction of {\em nerves} to substitute the
continuous topology.  Recall that the nerve $\kkk$ of a covering
$\uuu$ of a manifold $\mmm$ is the simplicial complex whose
vertices are the elements of $\uuu$ and whose simplices are formed
according to the following rule. A set of vertices (that is,
elements of the covering) $\{U_0, \ldots , U_k\}$ form a
$k$-simplex of $\kkk$ if and only if they have nonempty
intersection:

\[
\{U_0, \ldots , U_k\} \in \kkk \: \Leftrightarrow \:
U_0 \cap U_1 \cap \ldots \cap U_k \neq \emptyset
\]

\noindent Any nerve $\kkk$ being a simplex can be as well treated
as a poset, denoted also by $\kkk$. The points of the poset $\kkk$
are the simplices of the complex $\kkk$, and the arrows are drawn
according to the rule:

\[
p \to q \: \Leftrightarrow \: p \mbox{ is a face of } q
\]

\medskip

In the nondegenerate cases the posets associated with nerves and
those produced by Sorkin's (1991) `equivalence algorithm' are the
same. We choose nerves because their specific algebraic structure
makes it possible to build the dual algebraic theory.

\section{Incidence algebras}\label{s2}

The notion of incidence algebra of a poset was introduced by Rota
(1968) in a purely combinatorial context. Let $P$ be a poset.
Consider the set of formal symbols $\ketbra{p}{q}$ for all $p, q\in
P$ such that $p\le q$ and its linear span

\be{e38a}
\omg = \spn\{\ketbra{p}{q}\}_{p\le q}
\ee

\noindent and endow it with the operation of multiplication

\be{38}
\ketbra{p}{q} \cdot \ketbra{r}{s} =
\ket{p} \braket{q}{r} \bra{s} =
\braket{q}{r} \cdot \ketbra{p}{s} =
\left\lbrace \begin{array}{rcl}
\ketbra{p}{s} &,& \mbox{if } q=r \cr
0 && \mbox{otherwise}
\end{array} \right.
\ee

The correctness of this definition of the product, that is, the
existence of $\ketbra{p}{s}$ when $q=r$ is due to the transitivity
of the partial order. The obtained algebra $\omg$ with the
product (\ref{38}) is called {\em incidence algebra} of the poset
$P$.

The incidence algebra $\omg$ is obviously associative, but not
commutative in general. Namely, it is commutative if and only if
the poset $\ppp$ contains no arrows.

Let us split $\omg$ into two subspaces

\[
\omg = \aaa \oplus \rrr
\]

\noindent where

\be{e37a}
\aaa = \spn\{\ketbra{p}{p}\}_{p\in P}
\ee

\noindent and call

\[
\rrr = \spn \{ \ketbra{p}{q} \}_{p<q}
\]

\noindent the {\em module of differentials} of the poset $P$. It is
a fact that $\rrr$ is a bimodule over $\aaa$.

As we refine the poset, the limit space is intended to be a
manifold. The incidence algebras are dual objects to posets,
therefore their behavior should be similar to that of differential
forms in classical geometry. The algebra $\aaa$ is intended to play
the r\^ole of classical coordinates, while $\rrr$ should be graded
being an analogue of the module of differential forms.

For this aim we consider only simplicial complexes which are
treated as posets. $p\le q$ means that $p$ is a face of $q$. The
elements of simplicial comlexes are naturally graded. Then
any basic element $\ketbra{p}{q}$ of the incidence algebra $\omg$
acquires a degree being the difference of the degrees of its
constituents:

\be{e40a}
\degree\ketbra{p}{q} = \mbox{'the difference of cardinalities of
$p$ and $q$'}
\ee

\noindent splitting $\omg$ into linear subspaces

\be{40}
\omg = \omg^0 \oplus \omg^1 \oplus \ldots
\ee

\noindent with

\[
\begin{array}{rcl}
\omg^0 &=& \spn\{\ketbra{p}{p}\} \: = \: \aaa \cr
\ldots & \ldots & \ldots \cr
\omg^n &=& \spn\{\ketbra{p}{q}\}_{\deg\ketbra{p}{q} = n} \cr
\ldots & \ldots & \ldots
\end{array}
\]

\noindent making $\omg$ graded algebra:

\[
\forall \om \in \omg^m, \om' \in \omg^n \qquad
\om\om' \in \omg^{m+n}
\]

\noindent and therefore making the module of differentials
$\rrr$ graded $\aaa$-bimodule:

\[
\rrr = \omg^1 \oplus \omg^2 \oplus \ldots
\]

This grading makes the incidence algebras discrete differential
manifolds (Dimakis and M\"uller-Hoissen, 1999) as they posssess both
commutative scalars (the subalgebra $\aaa$) and differentials over it
(the module $\rrr$). For a more detailed account the reader is referred
to (Breslav and Zapatrin, 1999).

\section{Rota topology and the duality}\label{s2a}

In this section we establish a duality between a certain class of
finitary substitutes and their incidence algebras. We select
this class in such a way that canonical mappings between the
points admit conjugate homomorphisms of incidence algebras
making the correspondence between posets and algebras
functorial.

As it was shown in the previous section, with any poset its
non-commutative incidence algebra can be associated. It was proved by
Stanley (1968) that if two posets have isomorphic incidence algebras
then they are isomorphic. The reverse procedure building a poset
$P(\omg)$ from an arbitrary finite-dimensional algebra  $\omg$ was
suggested in (Zapatrin, 1998).  Let us briefly describe the
construction.

The elements of the poset $P(\omg)$ are the irreducible representations
of the algebra $\omg$. Building the partial order on $P(\omg)$ cosists
of two steps. First, the nearest neighbour connections $p\to q$ are
built according to the following rule: let $p,q$ are two irreducible
representations of $\omg$, denote by $p^0,q^0$ their kernels:

\[
p^0 = p^{-1}(0) \quad ; \quad q^0 = q^{-1}(0)
\]

\noindent which are ideals in $\omg$. Then define the nearest
neighbours $p\to q$:

\be{e93}
p\to q \quad \Leftrightarrow \quad
p^0q^0 \neq p^0\cap q^0
\ee

\noindent where the left-hand side $p^0q^0$ is understood as the
product of subsets of $\omg$. The resulting partial order on the set
$P(\omg)$ is obtained as the transitive closure of the relation
(\ref{e93}). The topology associated with this partial order is
referred to as {\em Rota topology}.

When the algebra $\omg$ is commutative, the Rota topology is discrete
(no linked pairs $p\to q$). The obtained topology becomes non-trivial
only when $\omg$ is non-commutative.

\paragraph{Remark.} When all irreducible representations of $\omg$ are
one-dimensional we can build two topologies on the set $P(\omg)$
Gel'fand and Rota ones, and it is interesting to compare them. The
result is the following: the Gel'fand topology is always discrete,
while the Rota topology may be non-trivial.

The possibility of mutual transitions between between finitary
topological spaces and algebras is based on the following theorem
(Zapatrin, 1998):

{\em
If the algebra $\omg$ is the incidence algebra $\omg(\ppp)$ of a poset
$\ppp$ then the resulting poset is isomorphic to $\ppp$:
}

\[
\ppp \simeq \ppp(\omg(\ppp))
\]

As it was mentioned, Stanley (1968) theorem claims that

\[
\omg(\ppp) \simeq \omg(\qqq)
\Leftrightarrow
\ppp \simeq \qqq
\]

\noindent and one could expect that a poset homomorphism, that
is, a continuous  mapping of appropriate finitary topological
spaces, should give rise to a homomorphism of their incidence
algebras. Alas, this is not the case for general posets $\ldots$

To gather functoriality we have to restrict somehow the class of
posets we are dealing with and the mappings between them. We did
it already in the previous section in order to make the
incidence algebras graded. Namely, we restricted ourselves to
simplicial complexes. To make incidence algebras dual objects,
we, following Alexandrov (1956), restrict the mappings between
simplicial complexes to simplicial mappings only. Recall that a
mapping $\om: \kkk_\alpha \to \kkk_{\alpha'}$ between two
simplicial complexes $\kkk_\alpha$ and $\kkk_{\alpha'}$ is said
to be simplicial if

\begin{itemize}
\item the $\om$-image of any vertex in $\kkk_\alpha$ is a vertex
in $\kkk_{\alpha'}$

\be{e77}
\om(\kkk^0_\alpha) \subseteq \kkk^0_{\alpha'}
\ee

\item $\om$ is completely defined by its values on the vertices
of $\kkk_\alpha$.

\item $\om$ preserves simplices
\end{itemize}

Under this condition the correspondence between posets and their
incidence algebras becomes functorial. With any $\om:
\kkk_\alpha \to \kkk_{\alpha'}$ its adjoint $\om^*:
\omg(\kkk_{\alpha'}) \to \omg(\kkk_\alpha)$ is defined in the
following way. Let $\ketbra{P}{Q}$ be a basic element of
$\omg(\kkk_{\alpha'})$. Then $Q$ may be represented as a
disjoint sum $Q = P+S$ and we put

\be{e79}
\om^*(\ketbra{P}{Q}) =
\sum\{ \ketbra{p}{q} \:: \om(p) = P;\, q=p+s,\, \om(s)=S \}
\ee

The direct verification shows that the so-defined $\om^*$ is
always a homomorphism of the incidence algebras. As a result, we
have the duality between simplicial complexes and their
incidence algebras: if $\om$ is surjective its adjoint $\om^*$
is injective and vice versa.

\section{Physical interpretation of incidence algebras}\label{s3}

The differential manifold model of spacetime is an ill-founded
assumption and a gross generalisation of what we actually
experience as spacetime. It is essentially based on the
non-operational supposition that we can pack an infinity of
events in an infinitesimal spacetime volume element when, in
fact, we only record a finite number of them during experiments
of finite duration in laboratories of finite size. It is exactly
due to this trait of the manifold model that at small scales our
theories of quantum spacetime structure and dynamics are plagued
by non-renormalizable infinities of the values of many important
physical fields. On the other hand, the requirement that the laws
of nature are local almost mandates the assumption of smoothness
for spacetime and we seem to get back to square one. However, the
success the manifold has enjoyed in picturing the local dynamics of
matter should not mask the unphysicality of its character,
especially at small scales. In particular, quantum theory, when
applied to investigate the structure and dynamics of spacetime in
the small, is simply incompatible with a classical,
non-operational ideal of a continuous infinity of events labelled
by commutative coordinates. Pragmatic measurements of quantum
spacetime are finite and inevitably induce uncontrollable
dynamical perturbations to it. Thus, the requirement for
operationality and finiteness as well as the success that a quantum
theory of matter has had when formulated algebraically motivate us
to formulate an algebraic theory of pragmatic finite measurements
of spacetime at quantum scales.

The local structure of the differential manifold is the point event
and its infinitesimal differential neighbours in the tangent space.
As mentioned above, this classical geometric structure serves us
well in casting dynamical laws as differential equations (classical
locality), but is rather inadequate for picturing actual quantum
spacetime measurements that are finite hence free from infinities
(singularities). Especially in the quantum deep this classical
conception of locality can not survive. We propose to revise it by
substituting the geometrical point events and the space of
directions tangent to them by finitely generated algebras affording
a cogent quantum spacetime interpretation for their structure. In
this sense our scheme is alocal and finitary and more likely to
evade the infinities of the differential manifold. Of course, the
'naturalness' of our substituting quantum alocality for the
classical differential locality will be vindicated if we are able
to recover the limit manifold with its classical observables and
differential structure by some kind of correspondence principle
applied to the alocal algebras of pragmatic measurements. We carry
this out in the section \ref{s4}.

We give the following physical meaning to the elements of $\omg$ in
(\ref{e37a}):

\begin{enumerate}
\item $\aaa$ constitutes the space of stationaries. The latter can
be thought of as elementary acts of determination of the pure
states of quantum spacetime. We interpret them as quantum
spacetime events. The algebraic connective '$+$' between them is
interpreted as coherent superposition between quantum events. The
commutativity of stationaries foreshadows the compatibility of the
determinations of the coordinates of events in the classical
manifold regime (Section \ref{s4}). In the dual (poset) picture the
sationaries correspond to self-incidences $p\to p$.

\item $\omg^1$ constitutes the space of transients. These can be
thought of as elementary quantum dynamical processes between
stationaries, thus they represent discrete one-step transitions
between quantum spacetime events. Transients do not commute with
each other and this foreshadows the Lie structure of covectors in
the limit space.

\item $\omg^i$ ($i\ge 2$) constitute the spaces of paths which are
thought of as composites of transients. If we associate with a
transient a quantum of an additive physical
quantity like energy (or its dual time), then the total grade of the
appropriate element of the algebra corresponds to the total energy
associated with it (or to the duration of the whole transition process).

\end{enumerate}

In the Motivation we alluded to the Sorkin poset scheme as being an
analog of the Schr\"odinger picture of quantum dynamics while our
algebraic approach as being the simile of the Heisenberg picture:
this is based on the duality of the two approaches (Section
\ref{s2a}). In an analogous way quantum states are the linear duals
of the operators in the conventional algebraic approach to quantum
mechanics.

Here too any finitary substitute is associated with an incidence
algebra in such a way that the topology of the poset is the same as
that encoded in the algebra. This resembles the fact that the
Schr\"odinger and the Heisenberg pictures encode the same
information about quantum dynamics.  Furthermore, the arrows
between point events in the Sorkin scheme can be thought of as the
directed dynamical transitions of spacetime event-states while in
our picture such dynamical connections are between pragmatic
operations. The topology in both schemes is physically interpreted
as dynamical connections between events although our picture being
algebraic naturally affords a quantum interpretation.

\section{Limiting procedure and the correspondence principle
}\label{s4}

When spacetimes are subsituted by finitary topological spaces,
we may consider finer or coarser experiments. That is why we
have to formalize the notion of refined experiment. Within the
Sorkin discretization procedure (Section \ref{s1}) a refinement
means passing to an inscribed covering of the manifold. In this
case any element of the finer covering is contained in an
element of the coarser one. Since we are dealing with nerves and
simplicial mappings between them we have to take care of the
condition (\ref{e77}). Recall that a vertex of the nerve is
associated with an element of the covering. In general it
may happen that a small region of the fine covering can belong
to two elements of a coarser one. So, we have have to require
for any element of the fine covering to keep track of its origin
in order for (\ref{e77}) to hold.

Each step of a limiting procedure, that is, a refined covering,
gives rise to a projection of appropriate complexes: the finer
one is projected to the coarser one. In the dual framework
we have an injection of the smaller algebra associated with a
coarser measurement to the bigger one.

In general, limiting procedures for approximating systems (whatever
they be, posets or algebras) are organised using the notion of
converging nets. Namely, each pragmatic observation is labelled by
an index $\alpha$ and we have the relation of refinement $\succ$ on
observations: $\alpha \succ \alpha'$ means that the observation
$\alpha$ is a refinement of $\alpha'$.

When we are dealing with posets with each pair $\alpha,\alpha'$
such that $\alpha \succ \alpha'$ a canonical projection
$\om_{\alpha'}^\alpha : \kkk_\alpha \to \kkk_{\alpha'}$ is defined
such that for any $\alpha \succ \alpha' \succ \alpha''$

\[
\om_{\alpha''}^\alpha =
\om_{\alpha''}^{\alpha'}\om_{\alpha'}^\alpha
\]

\noindent and we introduce the set of threads. A thread is a
collection $\{t_\alpha\}$ of elements $t_\alpha \in \kkk_\alpha$
such that

\[
t_{\alpha} = \om_{\alpha}^{\alpha'} t_{\alpha'}
\]

\noindent whenever $\alpha \succ \alpha'$. Denote by $\tttt$ the
set of all threads.

The next step is to make $\tttt$ a topological space which is done
in a standard way (Alexandrov, 1956): $\tttt$ is a subspace of the
total cartesian product

\[
\tttt_0 = \times_\alpha \kkk_\alpha
\]

\noindent while each of $\kkk_\alpha$ is a topological space.
Endow $\tttt_0$ with the product Tikhonov topology, then $\tttt$
being a subset of $\tttt_0$ becomes topological space.
Finally we obtain the limit space $X$ as the collection of all
closed threads from $\tttt$. This procedure is described in detail
in (Sorkin, 1991).

\medskip

The scheme for building limit algebras is exposed in (Landi and
Lizzi, 1998). As mentioned above, with any pair $\alpha \succ
\alpha'$ of pragmatic observations we have a canonical injection

\[
\om^{*\,\alpha'}_{\alpha} : \omg(\kkk_{\alpha'}) \to
\omg(\kkk_{\alpha})
\]

Moreover, due to the requirement (\ref{e77}) the restriction of
each $\om^*$ on commutative subalgebras $\aaa = \omg^0 \subseteq
\omg$ is well defined. Now we first consider the set of all
sequences

\[
{\bf \omg} = \times_\alpha \omg(\kkk_\alpha) =
\{ \{a_\alpha\}\,|\, a_\alpha \in \omg(\kkk_\alpha) \}
\]

\noindent and select the set of {\em converging} sequences in the
following way. Note that $\aaa$ is an algebra. Introduce a norm
$||\cdot||_\alpha$ in each finite-dimensional algebra
$\omg(\kkk_\alpha)$, then a sequence $\{a_\alpha\}$ converges if
and only if for any $\epsilon > 0$ there exists a filter
$\fff_\epsilon$ of indices $\alpha$ such that

\[
\forall \alpha, \alpha'\in \fff_\epsilon
\quad \alpha \succ \alpha' \Rightarrow
||\om^{*\,\alpha}_{\alpha'}a_\alpha - a_{\alpha'}||_{\alpha'} <
\epsilon
\]

Since any element of the limit algebra is a net we may consider the
coupling between the limit algebra and the limit space which
consists of nets. The result of this coupling is a converging net
of numbers whose limit is thought of as the value of an element of
the limit algebra at a point of the limit space.

The Sorkin scheme recovers the manifold in the limit of
refinements of finitary posets. Our dual picture aspires to the
same in the limit of resolution of pragmatic event
determinations. Since our algebraic scheme affords a quantum
spacetime interpretation, this limit can be thought of as a
correspondence principle linking the finitary quantum spacetime
substrata with the smooth classical spacetime manifold. The
alocal, algebraic quantum spacetime determinations of the substrata 
converge to the local geometric spacetime point and its cotangent 
space. This is to be contrasted for instance with the Bombelli {\em et 
al.} (1987) causal set scenario where the limiting procedure may be 
thought of as a `random sprinkling' of events according to some 
appropriate distribution so that the `limit spacetime manifold', with 
its topological, differential and Lorentz-causal structure, arises as a 
statistical average of causal sets, thus it is essentially of {\em 
thermodynamic} nature.

On the other hand, our correspondence limit is well-defined in the
{\em quantum} (rather than statistical) sense as the well-known
correspondence principle: the pragmatic quantum stationaries
decohere to the point events of the limit manifold, while the
non-commuting transients to covectors.

\section*{Concluding remarks}

In the present paper we gave quantum spacetime interpretation to
the incidence algebras induced by posets which, in turn,
correspond to finitary topological spaces. Sorkin's limit for
recovering the manifold as a maximal refinement of finitary
posets is cast here as Bohr's correspondence principle. Still,
due to the implausibility of any notion of pre-existing space in
the quantum dynamical deep, we would rather give a more
physical, causal or temporal interpretation to the posets'
partial order (Sorkin, 1995), so that we can link our algebraic scheme
with Bombelli {\em et al.} (1987) causal set approach to quantum
gravity. Our quantum interpretation of the incidence algebras
induced by causal sets is a first step into yet another attempt
at quantizing causality (Finkelstein, 1969). It is one of the
authors' previous result (Zapatrin, 1998) and Finkelstein's
(1985) claim for immediate causal links between
events to represent the physical causal topology that caught our
attention and motivated us to try to link the present work with causal
sets.  This project, however, is still at its birth.

\paragraph{Acknowledgments.} The work was carried out within the
program "T\^ete-\`a-t\^ete in St.Petersburg" under the auspices of the
Euler Mathematical Institute. The authors express their
gratitude to the participants of the Friedmann seminar
(St.Petersburg, Russia) for their kind suggestions.
One of the authors (IR) wholeheartedly acknowledges
numerous exchanges on quantum spacetime structures with Professor
Anastasios Mallios. The second author (RRZ) acknowledges the
support from the RFFI research grant and the research grant
"Universities of Russia".

\medskip

\noindent {\bf References}

Aleksandrov, P.S. (1956). 
{\it Combinatorial Topology}, 
Greylock, Rochester, New York
\smallskip

Bombelli L., Lee J., Meyer D., and Sorkin R.D. (1987).
Spacetime as a causal set,
{\it Physical Review Letters},
{\bf 59}, 521
\smallskip

Breslav, R., and R.R.Zapatrin (1999).
Differential structure of Greechie logics,
{\it International Journal of Theoretical Physics},
submitted, eprint quant-ph/9903011

Dimakis, A., and F.M\"uller-Hoissen (1999).
Discrete Riemannian geometry,
{\it Journal of Mathematical Physics},
{\bf 40}, 1518
\smallskip

Einstein, A. (1924). 
\"Uber den \"Ather,
{\it Schweizerische naturforschende Gesellschaft Verhanflungen},
{\bf 105}, 85-93 (English translation by Simon Saunders:
'On the Ether' in {\it The Philosophy of Vacuum},
S.Saunders and H.Brown, Eds.,
Oxford University Press (1991),
13--20)
\smallskip

Finkelstein D. (1969). 
Space-time code,
{\it Physical Review},
{\bf 184}, 1261
\smallskip

Finkelstein, D. (1985). 
Superconducting Causal Nets,
{\it International Journal of Theoretical Physics},
{\bf 27}, 473
\smallskip

Landi, G., and F.Lizzi (1999).
Projective Systems of Noncommutative Lattices as a Pregeometric
Substratum,
in {\it Quantum Groups and Fundamental Physical Applications},
ISI Guccia, Palermo, December 1997, D. Kastler and M. Rosso Eds.,
Nova Science Publishers, Inc.
\smallskip

Parfionov, G.N. and R.R.Zapatrin (1995).
Pointless Spaces in General Relativity,
{\it International Journal of Theoretical Physics},
{\bf 34},
737
\smallskip

Rota, G.-C., (1968).
On The Foundation Of Combinatorial Theory, I. The Theory Of
M\"obius Functions,
{\it Zetschrift f\"ur Wahrscheinlichkeitstheorie,\/}
{\bf 2},
340
\smallskip

Sorkin, R.D. (1991). 
Finitary Substitute for Continuous Topology,
{\it International Journal of Theoretical Physics}, {\bf 30}, 7, 923
\smallskip

Sorkin, R.D. (1995).
A specimen of theory construction from quantum gravity,
in {\it The Creation of Ideas in Physics},
Jarret Leplin Ed.,
Kluwer Academic Publishers,
Dordrecht
\smallskip

Stanley, R.P. (1986).
{\it Enumerative Combinatorics},
Wadsworth and Brook, Monterey, California
\smallskip

Zapatrin, R.R. (1998).
Finitary Algebraic Superspace,
{\it International Journal of Theoretical Physics},
{\bf 37},
799
\smallskip

\end{document}